\documentclass[superscriptaddress,twocolumn,amssymb,nofootinbib,10pt]{revtex4-2}
\usepackage[normalem]{ulem}
\usepackage[utf8]{inputenc}
\usepackage{comment}
\usepackage{graphicx}
\usepackage{graphics}
\usepackage{amsmath, amssymb}
\usepackage{graphics,bm}
\usepackage{graphicx}
\usepackage{bbold}
\usepackage{slashed}
\usepackage{wrapfig}
\usepackage{hyperref}
\usepackage{cancel}
\usepackage[usenames]{color}  
\usepackage{lipsum}
\usepackage[compat=1.1.0]{tikz-feynman}
\usetikzlibrary{arrows,shapes}
\usetikzlibrary{trees}
\usetikzlibrary{matrix,arrows} 			
\usetikzlibrary{positioning}				
\usetikzlibrary{calc,through}				
\usepackage{pgffor}							

\newcommand{\be}{\begin{equation}}
\newcommand{\ee}{\end{equation}}
\newcommand{\beq}{\begin{equation}}
\newcommand{\eeq}{\end{equation}}

\newcommand{\bqa}{\begin{eqnarray}}
\newcommand{\eqa}{\end{eqnarray}}

\newcommand{\ms}{\overline{\text{\tiny MS}}}

\def\square{\vcenter{\vbox{\hrule height.4pt
          \hbox{\vrule width.4pt height4pt
          \kern4pt\vrule width.3pt}\hrule height.4pt}}}

\usepackage[normalem]{ulem}

\begin{document}

\title{Renormalization of the three-flavor quark-meson diquark model}
\author{Jens O. Andersen}
\email{jens.andersen@ntnu.no}
\affiliation{Department of Physics, Faculty of Natural Sciences,NTNU, 
Norwegian University of Science and Technology, H{\o}gskoleringen 5,
N-7491 Trondheim, Norway}

\author{Mathias P. N{\o}dtvedt}
\email{mathias.p.nodtvedt@ntnu.no}

\affiliation{Department of Physics, Faculty of Natural Sciences,NTNU, 
Norwegian University of Science and Technology, H{\o}gskoleringen 5,
N-7491 Trondheim, Norway}

\date{\today}

\begin{abstract}
We discuss the properties of the two- and three-flavor quark-meson diquark (QMD) model as a renormalizable low-energy  effective model for color superconductivity in dense QCD. The effective degrees of freedom are scalars, pseudo-scalars, diquarks, and quarks. 
The parameters in the scalar/pseudo-scalar sector can be determined by matching
the meson pole masses and decay constants to their observed values using the on-shell renormalization scheme.
The remaining parameters are in the diquark sector and a priori unknown. In principle,
they can be calculated from QCD, but we consider them free. 
We renormalize the thermodynamic potential in the 2SC phase for two flavors
and in the color-flavor-locked (CFL) phase for three flavors, determining the counterterms of the couplings in the diquark sector.
We derive a set of renormalization group equations for these couplings that are used to improve the thermodynamic potential. As an application, we calculate the gap and the speed of sound in the ideal CFL phase.
It is shown that the gap approaches a constant as $\mu_B\rightarrow\infty$ and that 
the speed of sound relaxes to the conformal limit from above.

\end{abstract}

\maketitle
\noindent
\section{Introduction}
QCD at finite density has received a lot of attention in the decades following the discovery of asymptotic freedom at high energies. In the 1970s, it was thought that ultradense QCD matter consists of deconfined
quarks that interact weakly. Weak-coupling expansions of thermodynamic quantities at high densities have a long history dating back to this period~\cite{freed1,freed2,baluni}.
In recent years, there has been significant progress in perturbative calculations of
thermodynamic quantities. The weak-coupling expansion for the pressure
has been pushed to order $\alpha_s^3\log\alpha_s$, where
$\alpha_s$ is the strong coupling~\cite{dense1,dense4}. In Ref.~\cite{loic}, 
renormalization group techniques have been used to improve the convergence of perturbative series and reduce its scale dependence. Although perturbative QCD is not directly applicable to deconfined matter in compact stars, it has been used in conjunction with Bayesian methods to constrain the equation of state in a model-independent way~\cite{finnbayes}.

However, a complication of high-density QCD is that matter behaves
as a color superconductor~\cite{bcs1,bcs2,love}. There is an attractive quark-quark interaction arising from one-gluon exchange. It is known from BCS theory of normal superconductors that any attractive interaction
between fermions renders the Fermi surface unstable and Cooper pairs are formed.
For sufficiently large values of the baryon chemical potential, one can ignore the effects of the mass of the $s$-quark, and quarks with all flavors and colors participate in the pairing on an equal footing.
This phase is called the color-flavor-locked (CFL) phase and is the true ground state of QCD at asymptotically high baryon densities~\cite{alfordcolor,rappcolor,alfordcolor2}.
The CFL pairing gap has also been estimated using Bayesian methods
together with astrophysical observations~\cite{rachel}.
For lower values of the baryon chemical potential, one can no longer ignore the mass
of the strange quark. The mass of the $s$-quark and neutrality constraints put
stress on the CFL pairing. This gives rise to modified CFL phases involving 
kaon condensation, or a complete breaking of the pairs involving the $s$-quark.
In the latter case, only the $u$ and $d$ quarks pair up, giving rise
to the so-called 2SC phase.
Generally, many results are not robust, meaning that the position and even the
existence of certain phases in the phase diagram depend on the values of some couplings
in the effective models.
See Refs.~\cite{raja,alford,fukurev} for extensive reviews. 

Since lattice QCD is not available at large nonzero baryon chemical potentials due to the
sign problem, one strategy is to use low-energy models for QCD.
These models share some of the properties of QCD, for example, chiral symmetry breaking.
Conventionally, the quark phases of hybrid stars are described using the Nambu-Lona-Jasinio (NJL) model~\cite{baymrev}. The NJL model is a nonrenormalizable low-energy effective model whose ultraviolet divergences typically are regulated by a sharp three-dimensional cutoff  $\Lambda$. The NJL model suffers from regularization artifacts. For example, the behavior of diquark condensates is unphysical when the chemical potentials in the model are too high, i.e. close to the ultraviolet cutoff $\Lambda$. 
However, recent work~\cite{gholami} has improved the behavior of the model by introducing counterterms that depend on the chemical potentials.

In Refs.~\cite{us,us2}, we studied in detail the properties of the two-flavor quark-meson diquark (QMD) model as a
renormalizable low-energy effective model to describe the 2SC phase of dense QCD, See also Ref.~\cite{pawlow,braun22,bjs}.
The starting point is the quark-meson model (QM) which is used to describe normal quark matter. Describing the color superconducting phases requires the introduction of diquark effective degrees of freedom in the quark-meson model. The diquarks interact with mesons and quarks, and with themselves.
The addition of the diquarks gives rise to a few new terms in the low-energy effective Lagrangian. The allowed terms are dictated by symmetry and renormalizability. The couplings of these terms are a priori unknown, and we consider them free. In Refs.~\cite{us,us2}, it was found that the results are sensitive mainly to the diquark mass parameter $m_{\Delta}$ and the quark-diquark coupling $g_{\Delta}$.
By choosing $m_{\Delta}=500-900$ MeV and $g_{\Delta}\approx g$, where $g$ is the Yukawa coupling, we obtained a transition to 2SC matter at around $\mu_B=300$ MeV and a gap of 50-150 MeV in the range of baryon chemical potentials where the 2SC is believed to exist.
Finally, it was shown that in the limit $\mu\rightarrow\infty$, the gap approaches a constant and the speed of sound relaxes to the conformal limit from above.
In Refs.~\cite{us,us2}, the sigma and the three pions were included as
mesonic degrees of freedom. In this case, the model is equivalent to the $O(4)$
linear sigma model. This model does not allow for different light quark masses, and therefore no explicit isospin symmetry breaking. Even if the light quarks are degenerate in the vacuum, isospin can be broken at finite
density due to non-zero chemical potentials. For example, in the 2SC phase with charge neutrality imposed, the isospin chemical potential is non-vanishing, and might lead
to $m_u\neq m_d$. Therefore, we first consider the two-flavor system and show
how this feature can be incorporated into the model by extending the meson sector.

Of course, the two-flavor QMD model cannot describe the CFL phases since they involve
pairing of the $s$ quark too. In this paper, we extend the quark-meson diquark model to 
three flavors to be able to accommodate the CFL phases as well. 
We then define the three-flavor QMD model and show how the thermodynamic potential can be renormalized for arbitrary quark masses and CFL gaps.
As in the two-flavor case, a few new interaction terms between 
scalars and  diquarks and diquark self-interaction terms must be introduced.
Finally, we calculate the gap and the speed of sound at very large baryon chemical potentials, which show the same behavior as in the two-flavor case.

\section{Two-flavor quark-meson diquark model}
The properties of the two-flavor QMD model were discussed in some detail in~\cite{us}.
The degrees of freedom included are the sigma, pions, quarks, and diquarks.
The Minkowski-space Lagrangian is
\begin{widetext}
\begin{eqnarray}
\label{deflag2}
{\cal L}&=&{\cal L}_{\rm scalar}+{\cal L}_{\rm quark}+{\cal L}_{\rm diquark}
+{\cal L}_{\rm scalar-quark}
+{\cal L}_{\rm scalar-diquark}+{\cal L}_{\rm quark-diquark}\;.
\end{eqnarray}
where the different terms are
\begin{eqnarray}
\label{sss}
{\cal L}_{\rm scalar}&=&
\langle D_{\mu}\Sigma^{\dagger}D^{\mu}\Sigma\rangle-m^2\langle\Sigma^{\dagger}\Sigma\rangle-
{\lambda_1}\langle\Sigma^{\dagger}\Sigma\rangle^2
+\langle H(\Sigma+\Sigma^{\dagger})\rangle\;,
\\ 
{\cal L}_{\rm quark}&=&
\bar{\psi}(i\slashed{\partial}+\gamma^0\hat{\mu})\psi\;,
\\ 
{\cal L}_{\rm diquark}&=&
D_{\mu}\Delta^{\dagger}_aD^{\mu}\Delta_a
-m_{\Delta}^2\Delta^{\dagger}_a\Delta_a
-{\lambda_{\Delta}}(\Delta_a^{\dagger}\Delta_a)^2\;,
\\
{\cal L}_{\rm scalar-quark}&=&
-g\bar{\psi}(\sigma+i\gamma^5\vec{\tau}\cdot\vec{\pi})\psi\;,
\label{sq}
\\
{\cal L}_{\rm scalar-diquark}&=&-{\lambda_3}\langle\Sigma^{\dagger}\Sigma\rangle{\Delta}_a^{\dagger}\Delta_a\;,
\\
{\cal L}_{\rm quark-diquark}&=&
{1\over2}g_{\Delta}(\bar{\psi}^b)^C\Delta_a\gamma^5
\epsilon_{abc}\tau_2\psi^c
+{1\over2}g_{\Delta}\bar{\psi}^b\Delta_a^{\dagger}\gamma^5\epsilon_{abc}\tau_2
(\psi^c)^C\;. 
\label{lagrangian0}
\end{eqnarray}
\end{widetext}
Here $\langle A\rangle$ denotes the trace of the matrix $A$ and $C=i\gamma^0\gamma^2$
is the charge conjugation operator with $\psi^C=C\bar{\psi}^T$. 
The quark, meson and diquark fields are written as 
\begin{eqnarray}
\psi^T&=&(\psi_{u}^{r},\psi_{d}^{r},\psi_{u}^{g},\psi_{d}^{g},\psi_{u}^{b},\psi_{d}^{b})\;, 
\\
\Sigma&=&\sigma+i\vec{\tau}\cdot\vec{\pi}\;,\\
\Delta^T&=&(\Delta_1,\Delta_2,\Delta_3)\;.
\end{eqnarray}
The quarks $\psi_i^a$ have a color and a flavor index, as indicated.
The mesons are organized in a $2\times2$ matrix $\Sigma$ in terms of the unit matrix and the Pauli matrices $\tau_i$.
Under $SU(2)_L\times SU(2)_R$, the field $\Sigma$ transforms as $\Sigma\rightarrow L\Sigma R^{-1}$.
The diquark fields $\Delta_a$ have a color index $a$ and transform as an anti-triplet under $SU(3)_c$, and a singlet under $SU(2)_L\times SU(2)_R$.
Furthermore, $\hat{\mu}$ is the quark chemical potential matrix. The covariant derivative $D_{\mu}$ for the scalars and diquarks contains the appropriate chemical potentials.
They can be expressed in terms of the quark chemical potentials according to their
flavor and color quantum numbers.
The maximum number $N$ of chemical potentials that can be introduced simultaneously
is the same as the maximum of conserved charges in the system. In the present case,
these are baryon charge $B$, the electric charge $Q_e$, and the color charges $Q_3$ and 
$Q_8$. The quark chemical potentials $\mu_{ia}$ 
can then be expressed in terms of the four chemical 
potentials $\mu_B$,  $\mu_e$, $\mu_3$ and $\mu_8$ as
\begin{eqnarray}
\nonumber
\hat{\mu}_{ij,ab}&=&(\mbox{$1\over3$}\mu_B\delta_{ij}-\mu_eQ_{ij})\delta_{ab}
\\ &&
+\delta_{ij}\left[{1\over2}(T_3)_{ab}\mu_3+
{2\over\sqrt{3}}\mu_8
(T_8)_{ab}\right]\;,
\label{chempot}
\end{eqnarray}
where $i,j$ are flavor indices and $a,b$ are color indices.
Specifically, we find 
\begin{eqnarray}
\mu_{ur}&=&\mu-{2\over3}\mu_e+{1\over2}\mu_3+{1\over3}\mu_8\;,\\   
\mu_{dr}&=&\mu+{1\over3}\mu_e+{1\over2}\mu_3+{1\over3}\mu_8\;,\\   
\mu_{sr}&=&\mu+{1\over3}\mu_e+\frac{1}{2}\mu_3+{1\over3}\mu_8\;,\\
\mu_{ug}&=&\mu-{2\over3}\mu_e-{1\over2}\mu_3+{1\over3}\mu_8\;,\\
\mu_{dg}&=&\mu+{1\over3}\mu_e-{1\over2}\mu_3+{1\over3}\mu_8\;,\\
\mu_{sg}&=&\mu+{1\over3}\mu_e-\frac{1}{2}\mu_3+{1\over3}\mu_8\;,\\
\mu_{ub}&=&\mu-{2\over3}\mu_e-{2\over3}\mu_8\;,\\
\mu_{db}&=&\mu+{1\over3}\mu_e-{2\over3}\mu_8\;,\\
\mu_{sb}&=&\mu+{1\over3}\mu_e-{2\over3}\mu_8\;.
\end{eqnarray}
For later use, we also introduce the following average chemical potentials
\begin{eqnarray}
  \bar{\mu}_{ud}&=&   {1\over2}(\mu_{ur}+\mu_{dg}) ={1\over2}\left(\mu_{dr}+\mu_{ug}\right) \\
\bar{\mu}_{us}&=&
    {1\over2}\left(\mu_{ur}+\mu_{sb}\right) = {1\over2}\left(\mu_{ub}+\mu_{sr}\right) \\
\bar{\mu}_{ds} &=&    {1\over2}\left(\mu_{dg}+\mu_{sb}\right) = {1\over2}\left(\mu_{db}+\mu_{sg}\right) \;.    
\end{eqnarray}
Note that $\mu_3=0$ in two-flavor QCD~\cite{igor}, while it is generally nonzero in three-flavor QCD.

The explicit symmetry-breaking term $\langle H(\Sigma+\Sigma^{\dagger})\rangle=h_0\sigma$
does not allow for different quark masses since there is only one scalar field
that has a non-zero vacuum expectation value, namely $\sigma$.
In order to allow for isospin breaking, we must include the parity partners of $\sigma$ and $\vec{\pi}$, namely the pseudoscalar
$\eta$ and the scalar $\vec{a}$. This is done by generalizing the field $\Sigma$ to
\begin{eqnarray}
\Sigma&=&
T_a(\sigma_a+i\pi_a)=
{1\over2}(\sigma+i\eta)+{1\over2}\vec{\tau}\cdot(\vec{a}+i\vec{\pi})\;,    
\end{eqnarray}
where $T_0={1\over2}\mathbb{1}$, $T_i={1\over2}\tau_i$ for $i=1,2,3$ are the four generators of the $U(2)$ group. The quark-meson interaction term Eq.~(\ref{sq}) then reads
\begin{eqnarray}
\nonumber
{\cal L}_{\rm scalar-quark}&=&
-g\bar{\psi}T_0(\sigma+i\gamma^5\eta)\psi
-g\bar{\psi}\vec{T}\cdot(\vec{a}+i\gamma^5\vec{\pi})\psi\;,
\\ &&
\end{eqnarray}
We can express the Yukawa-like interaction in terms of left-handed and right-handed
fields as
\begin{eqnarray}
\bar{\psi}T_a(\sigma_a+i\gamma^5\pi_a)\psi&=&\bar{\psi}_L\Sigma\psi_R+
\bar{\psi}_R\Sigma^{\dagger}\psi_L\;.
\end{eqnarray}
In this form, the invariance under $SU(2)_L\times SU(2)_R$ transformations is evident.
Moreover, the coefficient of the symmetry breaking term is generalized to 
\begin{eqnarray}
H=T_0h_0+T_3h_3 \;.    
\end{eqnarray}
Denoting the expectation values of $\sigma$ and $a_3$ by $\bar{\sigma}$ and $\bar{a}_0$, respectively, ground state and the quark masses are
\begin{eqnarray}
\Sigma_0&=&T_0\bar{\sigma}_0+T_3\bar{a}_0=
{1\over2}
\left(\begin{array}{cc}
\bar{\sigma}_0+\bar{a}_0&0\\
0&\bar{\sigma}_0-\bar{a}_0\\
\end{array}\right)\;,
\\
m_u&=&{1\over2}g(\bar{\sigma}_0+\bar{a}_0)={1\over\sqrt{2}}g\phi_u\;, \\
m_d&=&{1\over2}g(\bar{\sigma}_0-\bar{a}_0)={1\over\sqrt{2}}g\phi_d\;.
\end{eqnarray}
In the scalar-pseudoscalar sector, this gives rise to new interaction terms in the Lagrangian 
\begin{eqnarray}
\nonumber
\delta{\cal L}_{\rm scalar}^{}&=&c\left[\det\Sigma+\det\Sigma^{\dagger}\right]
-{\lambda_2}\langle(\Sigma^{\dagger}\Sigma)^2\rangle
\\ 
\label{extra0}
&&-\lambda_{\rm det}\left[\det\Sigma+\det\Sigma^{\dagger}\right]
\langle\Sigma^{\dagger}\Sigma\rangle\;.
\end{eqnarray}
The determinant term is induced by instantons and is
invariant under $SU(2)_L\times SU(2)_R$. It equals $\sigma^2+\vec{\pi}^2-\eta^2-\vec{a}^2$.
Without the new fields, this term is equal to $\langle\Sigma^{\dagger}\Sigma\rangle$
and is therefore not an independent operator. 
There is also another independent quartic term, namely 
$\langle(\Sigma^{\dagger}\Sigma)^2\rangle=(\sigma^2+\vec{\pi}^2)^2+(\eta^2+\vec{a}^2)^2+(\sigma^2+\vec{\pi}^2)(\eta^2+\vec{a}^2)$. 
There is also a new interaction term between the scalars/pseudoscalars
and the diquarks, namely the product of the building blocks from each sector,
\begin{eqnarray}
\delta{\cal L}_{\rm scalar-diquark}&=&
-\lambda_{\rm det}^{\Delta}\left[\det\Sigma+\det\Sigma^{\dagger}\right]
\Delta_a^{\dagger}\Delta_a\;. 
\label{extra}
\end{eqnarray}
Eqs.~(\ref{deflag2}), ~(\ref{extra0}), and~(\ref{extra}) then define our 
extended two-flavor QMD model.
Since color superconductivity and pion condensation are mutually exclusive for two flavors~\cite{bedaque}, we can ignore the last term in Eq.~(\ref{extra0}) in the remainder.

\section{Renormalization of $\Omega$ in the two-flavor 2SC phase}
\label{2scsec}
We start by calculating the thermodynamic potential in the 2SC phase including 
leading quark-loop effects. The mesons are then treated at tree level.
The expectation value of the diquark anti-triplet is $(0,0,\Delta_{ud})$, where $\Delta_{ud}$ 
is the gap. Two of the three colors participate in the pairing, and the gap is antisymmetric in flavor and color. 
The ground state is invariant under chiral transformations
and the symmetry-breaking pattern is $SU(2)_L\times SU(2)_R\times SU(3)_c\rightarrow SU(2)_L\times SU(2)_R\times SU(2)_c$. Breaking of the global $SU(3)_c$
in the QMD model leads to the appearance of five Goldstone modes~\cite{sannino}. Thus, in QCD
there is no breaking of global symmetries and there is
no breaking of the $SU(3)_c$ gauge symmetry, but the Higgs mechanism leads to
five massive gluons.

Although the quasiparticle spectrum can be found for arbitrary values of the light quark masses and 
chemical potentials, the expressions are too complicated to be useful. Instead, we start from the expression for the quark-loop contribution to the thermodynamic potential.
The tree-level thermodynamic potential is
\begin{widetext}
\begin{eqnarray}
\nonumber
\Omega_0^{\rm 2SC}&=&{1\over2}m^2(\phi_u^2+\phi_d^2)
+(m_{\Delta}^2-4\bar{\mu}_{ud}^2)\Delta_{ud}^2
-h_u\phi_u-h_d\phi_d-c\phi_u\phi_d
+{1\over4}\lambda_1(\phi_u^2+\phi_d^2)^2
+{1\over4}\lambda_2(\phi_u^4+\phi_d^4)
\\ &&
+{1\over2}\lambda_3(\phi_u^2+\phi_d^2)\Delta_{ud}^2+{\lambda^{\Delta}}\Delta_{ud}^4
-\lambda_{\rm det}^{\Delta}\phi_u\phi_d\Delta_{ud}^2\;,
\end{eqnarray}
where $h_u={1\over\sqrt{2}}(h_0+h_3)$ and $h_d={1\over\sqrt{2}}(h_0-h_3)$.
The one-loop contributions to the thermodynamic potential from the quark loops is given by 
\begin{eqnarray}
\Omega_1^{\rm 2SC}&=&-{1\over2}\int_{p_0}\int_p\log\det S^{-1}\;,    
\label{smin1}
\end{eqnarray}
where $S^{-1}$ is the inverse quark propagator in the Gorkov basis,
\begin{eqnarray}
S^{-1}(p)&=&    
\left(\begin{array}{cc}
p\!\!\!/-\hat{m}+\gamma_0\hat{\mu}&ig_{\Delta}\tau_2\lambda_2\gamma^5\Delta_{ud}\\
ig_{\Delta}\tau_2\lambda_2\gamma^5\Delta_{ud}&p\!\!\!/-\hat{m}-\gamma_0\hat{\mu}\\
\end{array}\right)\;.
\end{eqnarray}
where \(\hat{m} = {\rm diag}(m_u,m_d)\) is the mass matrix. This is a $24\times24$ matrix, where the factor of ${1\over2}$ in Eq.~(\ref{smin1})
compensates for the doubling of the number of degrees of freedom.
The inverse propagator  in Nambu-Gorkov space, \(S^{-1}\), can be put in block diagonal form as described in \cite{RusterShovkovyRischke}.
The one-loop contribution to the thermodynamic potential from the quarks can then be 
written as
\begin{eqnarray}
\Omega_1^{\rm 2SC}&=&-\frac{1}{2}i\sum_{i=1,\pm}^3
\int_{-\infty}^{\infty}{dp_0\over2\pi}\int_p\log\det[p_0\mathbb{1}-M_i^{\pm}]\;,
\label{intdiv}
\end{eqnarray}
where the matrices are~\cite{matrices}
\begin{eqnarray}    
    M_1^{\pm} &=& \begin{pmatrix}
        m_u+\mu_{ug} & \pm p & 0 & \Delta_{ud}\\
        \pm p & -m_u+\mu_{ug} & -\Delta_{ud} & 0\\
        0 & -\Delta_{ud} & m_d-\mu_{dr} &\pm p\\
        \Delta_{ud} & 0 & \pm p & -m_d-\mu_{dr}
    \end{pmatrix}\;,
    \\
M_2^{\pm} &=& \begin{pmatrix}
        m_d+\mu_{dr} &\pm p & 0 & \Delta_{ud}\\
       \pm p & -m_d+\mu_{dr} & - \Delta_{ud} & 0\\
        0 & -\Delta_{ud} & m_u-\mu_{ug} & \pm p\\
        \Delta_{ud} & 0 &\pm p & -m_u-\mu_{ug}
    \end{pmatrix}\;,\\
M_{3}^{\pm}&=&\left(
\begin{array}{cccc}
-\mu_{db}+ m_d&\pm p&0&0\\
\pm p&-\mu_{db}- m_d&0&0\\
0&0&\mu_{ub}+ m_u&\pm p\\
0&0&\pm p&\mu_{ub}- m_u\\
\end{array}\right)\;.
\end{eqnarray}
The integral $\int_p$ is defined in $d=3-2\epsilon$ dimensions as
\begin{eqnarray}
\label{dimregdef}
\int_p&=&\left({e^{\gamma_E}\Lambda^2\over4\pi}\right)^{\epsilon}\int{d^dp\over(2\pi)^d}\;, 
\end{eqnarray}
with $\Lambda$ being the renormalization scale associated in the 
$\overline{\rm MS}$-scheme. In the calculations, we require two specific integrals given by 
\begin{eqnarray}
\label{i0}
\int_p\sqrt{p^2+m^2}&=&-{m^4\over2(4\pi)^2}\left({\Lambda\over m}\right)^{2\epsilon}
\left[{1\over\epsilon}+{3\over2}+{\cal O}(\epsilon)\right]
\;,\\
\int_p{1\over(p^2+m^2)^{3\over2}}&=&{4\over(4\pi)^2}\left({\Lambda\over m}\right)^{2\epsilon}
\left[{1\over\epsilon}+{\cal O}(\epsilon)\right]\;.
\label{i2}
\end{eqnarray}
The integrals in Eq.~(\ref{intdiv}) are divergent in four dimensions so we need to
regularize them. For arbitrary quark masses and gaps, the integral is too complicated
to be calculated directly in $d+1=4-2\epsilon$ dimensions.
The strategy is therefore as follows. We will construct subtraction terms that have the
same ultraviolet behavior as the integrand in Eq.~(\ref{intdiv}) and is sufficiently simple to be calculated using dimensional regularization. It can be constructed by expanding the integrand around vanishing chemical potentials and then integrating over $p_0$.
It is important that the subtraction term does not introduce any additional infrared divergences. The subtraction terms are given by 
\begin{eqnarray}
\nonumber
\Omega_1^{\rm 2SC, div}&=&-\Lambda^{-2\epsilon}\int_p\left[
2\sqrt{p^2+m_u^2}+2\sqrt{p^2+m_d^2}+
4\sqrt{p^2+m_u^2+g_{\Delta}^2\Delta_{ud}^2}+4\sqrt{p^2+m_d^2+g_{\Delta}^2\Delta_{ud}^2}
\right. \\ && \left.
-{(m_u-m_d)^2g_{\Delta}^2\Delta_{ud}^2\over(p^2+g_{\Delta}^2\Delta_{ud}^2)^{3\over2}}
+{(\mu_{ur}+\mu_{dg})^2g_{\Delta}^2\Delta_{ud}^2\over2(p^2+g_{\Delta}^2\Delta_{ud}^2)^{3\over2}}
+{(\mu_{ug}+\mu_{dr})^2g_{\Delta}^2\Delta_{ud}^2\over2(p^2+g_{\Delta}^2\Delta_{ud}^2)^{3\over2}}
\right]\;.    
\end{eqnarray}
The term
\begin{eqnarray}
\Omega_{\rm fin}^{\rm 2SC}&=&\Omega_1^{\rm 2SC}-\Omega_1^{\rm 2SC, div}\;,    
\end{eqnarray}
is by construction finite in $d=3$ dimensions and can be evaluated numerically.
Using the expressions for the integrals, we find
\begin{eqnarray}
\nonumber
\Lambda^{2\epsilon}\Omega_1^{\rm 2SC, div}&=&
{m^4_u\over(4\pi)^2}\left[{1\over\epsilon}+\log{\Lambda^2\over m_u^2}+{3\over2}\right]+
{m^4_d\over(4\pi)^2}\left[{1\over\epsilon}+\log{\Lambda^2\over m_d^2}+{3\over2}\right]+
{2(m^2_u+g_{\Delta}^2\Delta_{ud}^2)^2\over(4\pi)^2}\left[{1\over\epsilon}+\log{\Lambda^2\over m_u^2+g_{\Delta}^2\Delta_{ud}^2}+{3\over2}\right]
\\ &&
\nonumber
+{2(m^2_d+g_{\Delta}^2\Delta_{ud}^2)^2\over(4\pi)^2}\left[{1\over\epsilon}+\log{\Lambda^2\over m_d^2+g_{\Delta}^2\Delta_{ud}^2}+{3\over2}\right]
+{4(m_u-m_d)^2g_{\Delta}^2\Delta_{ud}^2\over(4\pi)^2}\left[{1\over\epsilon}+\log{\Lambda^2\over g_{\Delta}^2\Delta_{ud}^2}\right]
\\ &&
-{2(\mu_{ur}+\mu_{dg})^2\Delta_{ud}^2\over(4\pi)^2}\left[{1\over\epsilon}+\log{\Lambda^2\over g_{\Delta}^2\Delta_{ud}^2}\right]
-{2(\mu_{ug}+\mu_{dr})^2\Delta_{ud}^2\over(4\pi)^2}\left[{1\over\epsilon}+\log{\Lambda^2\over g_{\Delta}^2\Delta_{ud}^2}\right]\;. 
\label{om1vac}
\end{eqnarray}
In the $\overline{\rm MS}$-scheme, the wavefunction renormalization counterterms for the
scalars and pseudoscalars are the same and denoted by $\delta Z^{\ms}$.
The same remark applies to the parameters $h_u$ and $h_d$ as well
diquark fields $\Delta_a$, where the counterterms are
denoted by $\delta h^{\ms}$ and $\delta Z_{\Delta}^{\ms}$.
The counterterms are generated from the tree-level thermodynamical potential 
$\Omega_0^{{\rm 2SC}}$ and reads
\begin{eqnarray}
\nonumber
\Omega_1^{\rm 2SC, ct}&=&{1\over2}\delta m_{\ms}^2(\phi_u^2+\phi_d^2)
+{1\over2}m^2(\phi_u^2+\phi_d^2)\delta Z_{}^{\ms}
+\delta m_{\Delta}^2\Delta_{ud}^2
+(m_{\Delta}^2-4\bar{\mu}^2)\Delta_{ud}^2\delta Z_{\Delta}^{\ms}
-\delta h^{\ms}(\phi_u+\phi_d)
\\ && \nonumber
-{1\over2}(h_u\phi_u+h_d\phi_d)\delta Z_{\ms}
-\delta c_{\ms}\phi_u\phi_d-c\phi_u\phi_d\delta Z_{}^{\ms}
+{1\over4}(\delta\lambda_{1}^{\ms}+\delta\lambda_{2}^{\ms})(\phi_u^4+\phi_d^4)
+{1\over2}(\lambda_{1}+\lambda_2)(\phi_u^4+\phi_d^4)\delta Z_{\ms}
\\ &&
\nonumber
+{1\over2}\delta\lambda_{1}^{\ms}\phi_u^2\phi_d^2+{\lambda_1}\phi_u^2\phi_d^2\delta Z_{}^{\ms}
+{1\over2}{\delta\lambda_{3}^{\ms}}(\phi_u^2+\phi_d^2)\Delta_{ud}^2+{1\over2}\lambda_{3}(\phi_u^2+\phi_d^2)\Delta_{ud}^2(\delta Z^{\ms}+\delta Z_{\Delta}^{\ms})
\\ &&
+\delta\lambda_{\Delta}^{\ms}\Delta_{ud}^4+2\lambda_{\Delta}\Delta_{ud}^4\delta Z_{\Delta}^{\ms}-\delta\lambda_{\rm det}^{\Delta,\ms}\phi_u\phi_d\Delta_{ud}^2
-\lambda_{\rm det}^\Delta\phi_u\phi_d\Delta_{ud}^2(\delta Z^{\ms}+\delta Z_{\Delta}^{\ms})\;.
\label{ctexp}
\end{eqnarray}
\end{widetext}
In the $\overline{\rm MS}$-scheme, the poles in $\epsilon$ in 
$\Omega_1^{\rm 2SC,div}$, Eq.~(\ref{om1vac}), are exactly canceled by the 
counterterms from Eq.~(\ref{ctexp}). By comparing the various powers of $\phi_u$ and 
$\phi_d$, it is possible to directly determine most of the counterterms.
In other cases, we only obtain relations among counterterms. For example, we find   
\begin{eqnarray}
\label{rell1}
\delta m_{\ms}^2+m^2\delta Z_{}^{\ms}&=&0\;.
\end{eqnarray}
The counterterm $\delta Z^{\ms}$ can be determined by
calculating the two-point function for the meson field, including quark loops only.
This was done in Ref.~\cite{kumari,us} giving
\begin{eqnarray}
\delta Z_{}^{\ms}&=&-{3g^2\Lambda^{-2\epsilon}\over(4\pi)^2\epsilon}\;.    
\end{eqnarray}
Moreover, the counterterm for the Yukawa coupling $g$ is not found by the above matching. It is determined by noting that there is no renormalization of the quark-meson vertex or the quark fields to the approximation in which we are working. This yields the relation
$\delta g_{\ms}^2+\delta Z^{\ms}=0$.
The counterterms in the vacuum sector are
\begin{eqnarray}
    \delta m_{\ms}^2  &=&{3m^2g^2\Lambda^{-2\epsilon}\over(4\pi)^2\epsilon}\;,\\
    \delta g_{\ms}^2&=&{3g^2\Lambda^{-2\epsilon}\over(4\pi)^2\epsilon}\;,\\
    \delta c_{\ms}&=&{3g^2c\Lambda^{-3\epsilon}\over(4\pi)^2\epsilon}\;,\\
     \delta h^{\ms}&=&{3g^2h_u\Lambda^{-\epsilon}\over2(4\pi)^2\epsilon}\;,\\
    \delta Z_{}^{\ms}&=&-{3g^2\Lambda^{-2\epsilon}\over(4\pi)^2\epsilon}\;.\\
    \delta\lambda_1^{\ms}&=& \frac{6g^2\lambda_1\Lambda^{-4\epsilon}}{(4\pi)^2\epsilon}\\
    \delta\lambda_2^{\ms} &=& \frac{3g^2[2\lambda_2-g^2]\Lambda^{-4\epsilon}}{(4\pi)^2\epsilon}\;.
\end{eqnarray}
The counterterms involving the diquark fields, except $\delta g_{\Delta}^2$
can be determined by matching. Since there is no renormalization of the quark-diquark
vertex to the approximation in which we are working, we find the relation $\delta Z_{\Delta}^{\ms}+\delta g_{\Delta,\ms}^2=0$. The new counterterms are 
\begin{eqnarray}
    \delta m_{\Delta,\ms}^2  &=&{4g_{\Delta}^2\Lambda^{-2\epsilon}\over(4\pi)^2\epsilon}\;,\\
    \delta\lambda_{3}^{\ms}&=&{[3\lambda_3g^2+4\lambda_3g_{\Delta}^2-8g^2g_{\Delta}^2]\Lambda^{-4\epsilon}\over(4\pi)^2\epsilon}\;,\\
    \delta\lambda_{\rm det}^{\Delta,\ms}&=&{[3\lambda_{\det,\Delta}g^2+4\lambda_{\det,\Delta}g_{\Delta}^2-4g^2g_{\Delta}^2]\Lambda^{-4\epsilon}\over(4\pi)^2\epsilon}\;,\\
    \delta\lambda_{\Delta}^{\ms}&=&{4g_{\Delta}^2[2\lambda_{\Delta}-g_{\Delta}^2]\Lambda^{-4\epsilon}\over(4\pi)^2\epsilon}\;,\\
    \delta g_{\Delta,\ms}^2&=&{4g_{\Delta}^4\Lambda^{-2\epsilon}\over(4\pi)^2\epsilon}\;,\\
    \delta Z_{\Delta}^{\ms}&=&-{4g_{\Delta}^2\Lambda^{-2\epsilon}\over(4\pi)^2\epsilon}\;.\\
\end{eqnarray}
The bare parameters are independent of the renormalization scale $\Lambda$. This scale independence gives rise to a set of coupled differential equations for the running couplings. The solutions to these equations in the mesonic sector are
\begin{eqnarray}
    m_{\ms}^2(\Lambda)&=&{m_0^2\over[1-{3g_0^2\over(4\pi)^2}\log{\Lambda^2\over\Lambda_0^2}]}\;,\\
    \lambda_{1}^{\ms}(\Lambda)&=&{\lambda_{1,0}\over[1-{3g_0^2\over(4\pi)^2}\log{\Lambda^2\over\Lambda_0^2}]^2}\;,\\
    \lambda_{2}^{\ms}(\Lambda)&=&{\lambda_{2,0}-{3g_0^2\over(4\pi)^2}\log{\Lambda^2\over\Lambda_0^2}\over[1-{3g_0^2\over(4\pi)^2}\log{\Lambda^2\over\Lambda_0^2}]^2}\;,\\
 g_{\ms}^2(\Lambda)&=&{g_0^2\over[1-{3g_0^2\over(4\pi)^2}\log{\Lambda^2\over\Lambda_0^2}]}\;,\\
    c_{\ms}(\Lambda)&=&{c_0\over[1-{3g_0^2\over(4\pi)^2}\log{\Lambda^2\over\Lambda_0^2}]}\;,\\
    h_{f}^{\ms}(\Lambda)&=&{h_{f,0}\over[1-{3g_0^2\over(4\pi)^2}\log{\Lambda^2\over\Lambda_0^2}]^{1\over2}}\;,\\
    \phi_{f,\ms}^2(\Lambda)&=&\left[1-{3g_0^2\over(4\pi)^2}\log{\Lambda^2\over\Lambda_0^2}\right]\phi_{f,0}^2\;.
\end{eqnarray}
where $f=u,d$ and $m_0^2$, $g_0^2$...are the values of $m_{\ms}^2(\Lambda)$, $g_{\ms}^2(\Lambda)$... at the reference scale $\Lambda_0$.
Note that the products $m_{\ms}^2(\Lambda)\phi_{f,\ms}^2(\Lambda)$, 
$g_{\ms}^2(\Lambda)\phi_{f,\ms}^2(\Lambda)$, and $h_{f,\ms}(\Lambda)\phi_{f,\ms}(\Lambda)$ are scale-invariant quantities.
The solutions to the renormalization group equations in the diquark sector are
\begin{eqnarray}
    m_{\Delta,\ms}^2(\Lambda)&=&{m_{\Delta,0}^2\over\left[1-{4g_{\Delta,0}^2\over(4\pi)^2}\log{\Lambda^2\over\Lambda_0^2}\right]}\;,\\
    \lambda_{3}^{\ms}(\Lambda) &=& \frac{\lambda_{3,0}-\frac{8g_0^2g_{\Delta,0}^2}{(4\pi)^2}\frac{\Lambda^2}{\Lambda_0^2}}{\left[1-\frac{3g_{0}^2}{(4\pi)^2}\log\frac{\Lambda^2}{\Lambda_0^2}\right]\left[1-\frac{4g_{\Delta,0}^2}{(4\pi)^2}\log\frac{\Lambda^2}{\Lambda_0^2}\right]}\;,\\    
    \lambda_{\rm det,\Delta}^{\ms}(\Lambda)
    &=& \frac{\lambda_{\rm det,\Delta,0}-\frac{4g_0^2g_{\Delta,0}^2}{(4\pi)^2}\frac{\Lambda^2}{\Lambda_0^2}}{\left[1-\frac{3g_{0}^2}{(4\pi)^2}\log\frac{\Lambda^2}{\Lambda_0^2}\right]\left[1-\frac{4g_{\Delta,0}^2}{(4\pi)^2}\log\frac{\Lambda^2}{\Lambda_0^2}\right]}\;,\\    
 \lambda_{\Delta}^{\ms}(\Lambda)&=&{\lambda_{\Delta,0}-{4g_{\Delta,0}^4\over(4\pi)^2}\log{\Lambda^2\over\Lambda_0^2}\over\left[1-{4g_{\Delta,0}^2\over(4\pi)^2}\log{\Lambda^2\over\Lambda_0^2}\right]^2}\;,\\
     g_{\Delta,\ms}^2(\Lambda)&=&{g_{\Delta,0}^2\over\left[1-{4g_{\Delta,0}^2\over(4\pi)^2}\log{\Lambda^2\over\Lambda_0^2}\right]}\;,\\
    \Delta^2_{ud,\ms}(\Lambda)&=&\left[1-{4g_{\Delta,0}^2\over(4\pi)^2}\log{\Lambda^2\over\Lambda_0^2}\right]\Delta_{ud,0}^2\;,
\end{eqnarray}
where $m_{\Delta,0}^2$, $g_{\Delta,0}^2$... are the values of the running parameters at the reference scale $\Lambda_0$.
We notice that the products $m^2_{\ms}(\Lambda)g^2_{\Delta,\ms}(\Lambda)$
and $\Delta^2_{\ms}(\Lambda)g^2_{\Delta,\ms}(\Lambda)$ are renormalization group invariant.
After renormalization and inserting the expressions for the running parameters, we obtain
the thermodynamic potential in the 2SC phase
\begin{widetext}

\begin{eqnarray}
    \Omega^{\rm 2SC}_{0+1} &=& \Omega_0^{\rm 2SC} 
    -{16g_{\Delta,0}^2\bar{\mu}_{ud}^2\Delta_{ud,0}^2\over(4\pi)^2}\log{\Lambda_0^2\over g_{\Delta,0}^2\Delta_{ud,0}^2}
    + \frac{g_0^4\phi_{u,0}^4}{4(4\pi)^2}\left[\log{\Lambda_0^2\over m_u^2}+2\log{\Lambda_0^2\over m_u^2+g_{\Delta,0}^2\Delta_{ud,0}^2}+\frac{9}{2}\right]+ u\rightarrow d\nonumber\\
    && +\frac{2g_0^2g_{\Delta,0}^2\phi_{u,0}^2\Delta_{ud,0}^2}{(4\pi)^2}\left[\log{\Lambda_0^2\over m_u^2+g_{\Delta,0}^2\Delta_{ud,0}^2}+\log{\Lambda_0^2\over g_{\Delta,0}^2\Delta_{ud,0}^2}+3\right]+ u\rightarrow d\nonumber\\
    && +\frac{2g_{\Delta,0}^4\Delta_{ud,0}^4}{(4\pi)^2}\left[\log{\Lambda_0^2\over m_u^2+g_{\Delta,0}^2\Delta_{ud,0}^2}+\log{\Lambda_0^2\over m_d^2+g_{\Delta,0}^2\Delta_{ud,0}^2}+3\right]\nonumber\\
    &&-\frac{4g_0^2g_{\Delta,0}^2\phi_{u,0}\phi_{d,0}\Delta_{ud,0}^2} {(4\pi)^2}\log\frac{\Lambda_0^2}{g_{\Delta,0}^2\Delta_{ud,0}^2}+ \Omega_1^{\rm 2SC, fin}\;.
\end{eqnarray}
\end{widetext}
\section{Three-flavor quark-meson diquark model}
In this section, we present the three-flavor quark-meson diquark model and discuss its
symmetries. The effective degrees of freedom are scalars, pseudo-scalars, quarks,
and diquarks. The scalar sector was discussed in detail in Refs.~\cite{lenaghan,roder}.
The scalar fields are denoted by $\sigma_a$ and the pseudo-scalars by
$\pi_a$. They are organized in the matrix
\begin{eqnarray}
\Sigma&=&T_a(\sigma_a+i\pi_a)\;,
\end{eqnarray}
where $T_a$ are the generators of $U(3)$ with $T_0=\sqrt{{1\over6}}\mathbb{1}$
and $\lambda_a=2T_a$ are 
Gell-Mann matrices ($a=1,2,3...8$). The Gell-Mann matrices satisfy $\langle\lambda_a\lambda_b\rangle=2\delta_{ab}$, where $\langle A\rangle$ is the
trace of the matrix $A$.
The quarks are denoted by $\psi_i^a$, where the superscript $a$ ($a=1,2,3$)
is the color index, while the subscript $i$ ($i=1,2,3$) is the flavor index.
The quarks are organized in a color triplet and a flavor triplet $\psi$ as
$\psi^T=(\psi_{u}^{r},\psi_{d}^{r},\psi_{u}^{g},\psi_{d}^{g},\psi_{u}^{b},\psi_{d}^{b})$.
Finally, the diquarks
$\Delta_i^a\sim\psi^b_j\epsilon_{abc}\epsilon_{ijk}\gamma^5(\psi_k^c)^C$ 
are divided into left-handed and right-handed fields $\Delta_{L,i}^a$ and $\Delta_{R,i}^a$ with a color index $a$ and a flavor index $i$. They are organized into two
unitary $3\times3$ matrices that transform as $\Delta_L\rightarrow L\Delta_L U_c^T$ and $\Delta_R\rightarrow R\Delta_L U_c^T$, where $U_c$ is an $SU(3)_c$ transformation~\cite{sonstep,alfordlr}.

The Minkowski-space Lagrangian is
\begin{widetext}
\begin{eqnarray}
{\cal L}&=&{\cal L}_{\rm scalar}+{\cal L}_{\rm quark}
+{\cal L}_{\rm diquark}
+{\cal L}_{\rm scalar-quark}
+{\cal L}_{\rm scalar-diquark}
+{\cal L}_{\rm quark-diquark}\;,
\end{eqnarray}
where the different terms are
\begin{eqnarray}
{\cal L}_{\rm scalar}&=&\langle D_{\mu}\Sigma^{\dagger}D^{\mu}\Sigma\rangle-m^2\langle\Sigma^{\dagger}\Sigma\rangle-{\lambda_1}\langle\Sigma^{\dagger}\Sigma\rangle^2
-{\lambda_2}\langle(\Sigma^{\dagger}\Sigma)^2\rangle+\langle H(\Sigma+\Sigma^{\dagger})\rangle
+c[\det\Sigma+\det\Sigma^{\dagger}]\;,
\\ 
{\cal L}_{\rm quark}&=&\bar{\psi}(i\slashed{\partial}+\gamma^0\hat{\mu})\psi\;,
\\
\nonumber
{\cal L}_{\rm diquark}&=&
D_{\mu}(\Delta_{L,i}^a)^{\dagger}D^{\mu}\Delta_{L,i}^a
-m_{\Delta}^2(\Delta_{L,i}^a)^{\dagger}\Delta_{L,i}^a
-\lambda^{\Delta}_1[(\Delta_{L,i}^a)^{\dagger}\Delta_{L,i}^a]^2
-\lambda^{\Delta}_2(\Delta_{L,i}^a)^{\dagger}\Delta_{L,k}^a
(\Delta_{L,k}^b)^{\dagger}\Delta_{L,i}^b
\;,
\\ &&+{L\rightarrow R} -\lambda_3^{\Delta}
[(\Delta_{L,i}^a)^{\dagger}\Delta_{L,i}^a][(\Delta_{R,j}^b)^{\dagger}\Delta_{R,j}^b]
\\
{\cal L}_{\rm scalar-quark}&=&-g\bar{\psi}T_a(\sigma_a+i\gamma^5{\pi}_a)\psi\;,
\\
\nonumber
{\cal L}_{\rm scalar-diquark}&=&
-{\lambda_3}\langle\Sigma^{\dagger}\Sigma\rangle(\Delta_{L,i}^a)^{\dagger}\Delta_{L,i}^a
-\lambda_4\Sigma_{ij}^{\dagger}\Delta_{L,k}^a(\Delta^{a}_{L,j})^{\dagger}\Sigma_{ki}+{L\rightarrow R}\;,
\\ &&+\lambda_5\Sigma_{ij}(\epsilon_{klj}\Delta_{R,k}^ a)^\dagger\Sigma_{ml}\epsilon_{nmi}\Delta_{L,n}^a+\lambda_5\Sigma_{ij}^\dagger(\epsilon_{klj}\Delta_{L,k}^ a)^\dagger\Sigma_{lm}^\dagger\epsilon_{nmi}\Delta_{R,n}^a
\\ 
{\cal L}_{\rm quark-diquark}&=&
-{1\over2\sqrt{2}}g_{\Delta}(\bar{\psi}^b_{L,j})^C\Delta^a_{L,i}\gamma^5
\epsilon_{abc}\epsilon_{ijk}\psi_{L,k}^c
-{1\over2\sqrt{2}}g_{\Delta}\bar{\psi}_{L,j}^b(\Delta^a_{L,i})^{\dagger}\gamma^5\epsilon_{abc}\epsilon_{ijk}
(\psi^c_{L,k})^C-{L\rightarrow R}
\;,
\label{lagrangian}
\end{eqnarray}
\end{widetext}
where $H$ is the coefficient of the symmetry breaking term given by
\begin{eqnarray}
H&=&T_ah_a\;.
\end{eqnarray}
The determinant term arises from instantons and explicitly breaks the $U(1)_A$ symmetry.
The covariant derivatives of the mesons and diquarks contain the appropriate chemical potentials that reflect the conserved quantum numbers.

Under $SU(3)_L\times SU(3)_R$ transformations, the field $\Sigma$ transforms as
$\Sigma\rightarrow L\Sigma R^{-1}$.
The fields $\Sigma$ are organized in a $3\times3$
matrix with entries denoted by $\Sigma_{ij}$ where
the first index $i$ refers to left-handed transformations and the second index $j$ to
right-handed transformations.~\footnote{The indices of $\Sigma_{ij}^{\dagger}$
is then "right-left".}
Thus, the term $\langle\Sigma^{\dagger}\Sigma\rangle$, and powers thereof, are invariant.
We note in passing that for $N_f=3$, we have not included interaction terms involving
$[\det\Sigma+\det\Sigma^{\dagger}]$ since they are dimension five operators
and therefore irrelevant.

Next, we consider terms that involve diquark fields $\Delta_{L,i}^a$. Polynomial terms are constructed by saturating flavor and color indices. The mass term is then $(\Delta_{L,i}^a)^{\dagger}\Delta_{L,i}^a$. Regarding terms that involve four diquark fields, this can be done in two ways, either by combining two identical terms $(\Delta_{L,i}^a)^{\dagger}\Delta_{L,i}^a$, or by writing $(\Delta_{L,i}^a)^{\dagger}\Delta_{L,j}^b(\Delta_{L,j}^b)^{\dagger}\Delta_{L,i}^a$. A term that directly couples left-handed diquarks to right handed-diquarks is also allowed for symmetry reasons, with a separate coupling 
\(\lambda_3^\Delta\).
Now, consider quartic terms built from two meson fields and two diquark fields.
An obvious operator is the product, that is,
$\langle\Sigma^{\dagger}\Sigma\rangle(\Delta_{L,i}^a)^{\dagger}\Delta_{L,i}^a$.
The second operator is constructed from $\Sigma^{\dagger}_{ij}\Sigma_{kl}$ 
$(\Delta_{L,m}^a)^{\dagger}\Delta_{L,n}^b$, where we contract the color index setting $a=b$ and then the flavor indices, setting $i=l$, $j=n$ and $k=m$. This is the term propertional to $\lambda_4$ in Eq.~(\ref{lagrangian}). The same terms involving right-handed fields can be constructed.
There is a determinant-like term $\epsilon_{ijk}\epsilon_{lmn}\Sigma^{\dagger}_{li}\Sigma^\dagger_{jm}(\Delta_{L,k}^a)^{\dagger}\Delta_{R,n}^a$, combining left-handed and right-handed diquarks~\cite{su3}.
Finally, we consider the quark-diquark term. It is of the form $\bar{\psi}_i^b\Delta_j^a\psi^c_k$. Contracting it with the tensors $\epsilon_{ijk}$ and $\epsilon_{abc}$ in flavor and color space, respectively, we obtain a totally antisymmetric term~\cite{igor}.  The ground state is
\begin{eqnarray}
\Sigma_0&=&T_0\bar{\sigma}_0+T_3\bar{\sigma}_3+T_8\bar{\sigma}_8\;,   
\end{eqnarray}
where $\bar{\sigma}_0$ and $\bar{\sigma}_8$ are the expectation values of $\sigma_0$ and $\sigma_8$. The first term breaks the symmetry down to $SU(3)_V$ and the second term to $SU(2)_2\times SU(2)_R$ so that the vacuum state $\Phi_0$ is invariant under $SU(2)_V$.
Instead of using the basis $T_0$, $T_3$ and $T_8$, we introduce the linear combinations
\begin{eqnarray}
\phi_u&=&\sqrt{{2\over3}}\bar{\sigma}_0+\bar{\sigma}_3+\sqrt{{1\over3}}\bar{\sigma}_8\;,\\
\phi_d&=&\sqrt{{2\over3}}\bar{\sigma}_0-\bar{\sigma}_3+\sqrt{{1\over3}}\bar{\sigma}_8\;,\\
\phi_s&=&\sqrt{{1\over3}}\bar{\sigma}_0-\sqrt{{2\over3}}\bar{\sigma}_8\;.
\end{eqnarray}
In this basis, the ground state and the quark masses are
\begin{eqnarray}
\Sigma_0&=&{1\over2}{\rm diag}(\phi_u,\phi_d,\sqrt{2}\phi_s)\;,
\\
m_{u}&=&{1\over2}g\phi_u\;,\\
m_d&=&{1\over2}g\phi_d\;,\\
m_s&=&{1\over\sqrt{2}}g\phi_s\;.
\end{eqnarray}\\
\section{Renormalization of $\Omega$ in the three-flavor CFL phase}
Next, we consider the color-flavor-locked phases of dense QCD. 
In these phases, quarks of all flavors and colors pair up. The right-handed and 
left-handed condensates are equal up to a sign, and the expectation value of the diquark
matrix is $\Delta_L = \frac{1}{\sqrt{2}}{\rm diag}(\Delta_{us},\Delta_{ds},\Delta_{ud})$.
In the limit of three massless quarks, the breaking of the symmetry by the condensate is 
$SU(3)_L\times SU(3)_R\times SU(3)_c\times U(1)_B\times U(1)_A\rightarrow SU(3)_{L+R+c}$. 
In QCD, this gives rise to 10 Goldstone bosons, as well as 8 massive gluons via
the Higgs mechanism.
In the QMD model, the broken symmetries are global, leading to 18 massless
excitations. 

In the CFL phase with different quark masses, different gaps, and different quark chemical potentials, the dispersion relations are even more complicated than in the two-flavor case. Again, we will expand the quark determinants around vanishing chemical potentials. The resulting expressions are integrated over
the zeroth component of the four-momentum $p_0$. In this way, we isolate the ultraviolet behavior of the quark determinants. The terms generated by the expansion are added and subtracted. The added terms are evaluated using dimensional regularization and the remainder is finite in the ultraviolet and can be
calculated numerically directly in three dimensions. 
The tree-level thermodynamic potential is
\begin{widetext}
\begin{eqnarray}
    \Omega_0^{\rm CFL} &=& {1\over4}m^2(\phi_u^2+\phi_d^2+2\phi_s^2)-h_u\phi_u-h_d\phi_d-\sqrt{2}h_s\phi_s + \frac{1}{16}\lambda_1(\phi_u^2+\phi_d^2+2\phi_s^2)^2 + \frac{1}{16}\lambda_2(\phi_u^4+\phi_d^4+4\phi_s^4)\nonumber\\
    &&-\frac{c}{2\sqrt{2}}\phi_u\phi_d\phi_s+\frac{1}{4}\lambda_3(\phi_u^2+\phi_d^2+2\phi_s^2)(\Delta_{ud}^2+\Delta_{us}^2+\Delta_{ds}^2) +\frac{1}{4}\lambda_4[\phi_u^2\Delta_{ds}^2+\phi_d^2\Delta_{us}^2 +2\phi_s^2\Delta_{ud}^2]\nonumber\\
    && -\frac{1}{2}\lambda_{5}(\phi_u\phi_d\Delta_{ud}^2 + \sqrt{2}\phi_u\phi_s\Delta_{us}^2+\sqrt{2}\phi_d\phi_s\Delta_{ds}^2)\nonumber +(m_\Delta^2 - 4\bar{\mu}_{ud}^2)\Delta_{ud}^2 +(m_\Delta^2 - 4\bar{\mu}_{us}^2)\Delta_{us}^2
    \\ &&
    + (m_\Delta^2 - 4\bar{\mu}_{ds}^2)\Delta_{ds}^2 + \frac{1}{4}\left(2\lambda^{\Delta}_{1}+\lambda_3^\Delta\right)(\Delta_{ud}^2+\Delta_{us}^2+\Delta_{ds}^2)^2 + \frac{1}{2}\lambda^{\Delta}_{2}(\Delta_{ud}^4+\Delta_{us}^4+\Delta_{ds}^4)\;,
\end{eqnarray}

where we have defined
\begin{eqnarray}
h_u&=&\left({1\over\sqrt{6}}h_0+{1\over2}h_3+{1\over2\sqrt{3}}h_8\right)\;,\\    
h_d&=&\left({1\over\sqrt{6}}h_0-{1\over2}h_3+{1\over2\sqrt{3}}h_8\right)\;,\\    
h_s&=&\left({1\over\sqrt{6}}h_0-{1\over\sqrt{3}}h_8\right)\;.  
\end{eqnarray}
In analogy with the two-flavor case, the one-loop contribution to the thermodynamic potential from the quarks can be written as
\begin{eqnarray}
\Omega_1^{\rm CFL}&=&
-\frac{1}{2}i\sum_{i=1,\pm}^7
\int_{-\infty}^{\infty}{dp_0\over2\pi}\int_p\log\det[p_0\mathbb{1}-M_i^{\pm}]\;,
\end{eqnarray}
where the matrices $M_1^{\pm}$--$M_6^{\pm}$ are
\begin{eqnarray}
    M_1^{\pm} &=& \begin{pmatrix}
        m_u+\mu_{ug} & \pm p & 0 & \Delta_{ud}\\
       \pm p & -m_u+\mu_{ug} & -\Delta_{ud} & 0\\
        0 & -\Delta_{ud} & m_d-\mu_{dr} &\pm p\\
        \Delta_{ud} & 0 &\pm p & -m_d-\mu_{dr}
    \end{pmatrix}\;,\\
    M_2^{\pm} &=& \begin{pmatrix}
        m_d+\mu_{dr} &\pm p & 0 & \Delta_{ud}\\
        \pm p & -m_d+\mu_{dr} & - \Delta_{ud} & 0\\
        0 & -\Delta_{ud} & m_u-\mu_{ug} &\pm  p\\
        \Delta_{ud} & 0 &\pm p & -m_u-\mu_{ug}
    \end{pmatrix}\;,\\
    M_3^{\pm} &=& \begin{pmatrix}
        m_u-\mu_{ub} &\pm p & 0 & -\Delta_{us}\\
        \pm p & -m_u-\mu_{ub} & \Delta_{us} & 0\\
        0 & \Delta_{us} & m_s +\mu_{sr} &\pm p\\
        -\Delta_{us} & 0 &\pm p & -m_s+\mu_{sr}
    \end{pmatrix}\;,\\
    M_4^{\pm} &=& \begin{pmatrix}
        m_s-\mu_{sr} &\pm p & 0 & -\Delta_{us}\\
        \pm p & -m_s-\mu_{sr} & \Delta_{us} & 0\\
        0 & \Delta_{us} & m_u+\mu_{ub} &\pm p\\
        -\Delta_{us} & 0 &\pm p & -m_u+\mu_{ub}
    \end{pmatrix}\;,\\
        M_5^{\pm} &=& \begin{pmatrix}
        m_d-\mu_{db} &\pm p & 0 & -\Delta_{ds}\\
       \pm p & -m_d - \mu_{db} & \Delta_{ds} & 0\\
        0 & \Delta_{ds} & m_s+\mu_{sg} &\pm p\\
        -\Delta_{ds} & 0 &\pm p & -m_s+\mu_{sg}
    \end{pmatrix}\;,\\
    M_{6}^{\pm} &=& \begin{pmatrix}
    m_s - \mu_{sg} &\pm p & 0 & -\Delta_{ds}\\
    \pm p & -m_s-\mu_{sg} & \Delta_{ds} & 0\\
    0 & \Delta_{ds} & m_d+\mu_{db} &\pm p\\
    -\Delta_{ds} & 0 &\pm p & -m_d+\mu_{db}
    \end{pmatrix}\;.
\end{eqnarray}
The seventh matrix, $M_7^{\pm}$, is a $12\times12$ matrix that takes the form
\begin{eqnarray}
    M_7^{\pm} &=&\begin{pmatrix}
    M_7^{ur} & M_7^{\Delta_{ud}} & M_7^{\Delta_{us}}\\
    M_7^{\Delta_{ud}} & M_7^{dg} & M_7^{\Delta_{ds}}\\
    M_7^{\Delta_{us}} & M_7^{\Delta_{ds}} & M_7^{sb}
    \end{pmatrix}\;,
\end{eqnarray}
where the submatrices are defined as
\begin{eqnarray}
    M_7^{fc\pm} &=& \begin{pmatrix}
        -\mu_{fc}-m_f & \pm p & 0 & 0\\
       \pm p & -\mu_{fc}+m_f  & 0 & 0\\
        0 & 0 & \mu_{fc}-m_f &\pm p\\
        0 & 0 &\pm p & \mu_{fc}+m_f
    \end{pmatrix}\;,
    \hspace{1mm}
    M_7^{\Delta_{f_1f_2}} = \begin{pmatrix}
        0 & 0 & 0 & -\Delta_{f_1f_2}\\
        0 & 0 & \Delta_{f_1f_2} & 0\\
        0 & \Delta_{f_1f_2} & 0 & 0\\
        -\Delta_{f_1f_2} & 0 & 0 & 0
    \end{pmatrix}\;,
\end{eqnarray}
and where $f_1$ and $f_2$ are flavor indices. 

\normalsize
It turns out that the divergence structure is similar to the 2SC phase 
discussed in Sec.~\ref{2scsec}. The divergent subtraction term is
\begin{eqnarray}
    \Omega_1^{\rm CFL, div}&=&
\nonumber
-\Lambda^{-2\epsilon}\int_p\left\{
4\sqrt{p^2+m_u^2+g_{\Delta}^2\Delta_{ud}^2}+4\sqrt{p^2+m_u^2+g_{\Delta}^2\Delta_{us}^2} -2\sqrt{p^2+m_u^2} + {\rm cyclic\,\, perm.\,\,in\,\, flavor}
\right. \\ && 
\nonumber
\left.
-{(m_u-m_d)^2g_{\Delta}^2\Delta_{ud}^2\over(p^2+g_{\Delta}^2\Delta_{ud}^2)^{3\over2}}
+{(\mu_{ur}+\mu_{dg})^2g_{\Delta}^2\Delta_{ud}^2\over2(p^2+g_{\Delta}^2\Delta_{ud}^2)^{3\over2}}
+{(\mu_{ug}+\mu_{dr})^2g_{\Delta}^2\Delta_{ud}^2\over2(p^2+g_{\Delta}^2\Delta_{ud}^2)^{3\over2}}+{\rm cyclic\,\,perm.\,\,of\,\,color\,\,and\,\,flavor}
\right.\\ &&\left.
-{g_{\Delta}^4(\Delta_{ud}^2\Delta_{us}^2+\Delta_{ud}^2\Delta_{ds}^2+\Delta_{us}^2\Delta_{ds}^2)\over(p^2+M^2)^{3\over2}}
\right\}\;,
\end{eqnarray}
where $M$ is an arbitrary mass. The $M$-dependence drops out in the final result.
Performing the integrals using equations \eqref{i0} and \eqref{i2}, we find
\begin{eqnarray}
    \Omega_1^{\rm CFL, div}&=& \Lambda^{-2\epsilon}\left\{\frac{2(m_u^2+g_\Delta^2\Delta_{ud}^2)^2}{(4\pi)^2}\left(\frac{1}{\epsilon}+\frac{3}{2}+\log\frac{\Lambda^2}{m_u^2+g_\Delta^2\Delta_{ud}^2}\right) + \frac{2(m_u^2+g_\Delta^2\Delta_{us}^2)^2}{(4\pi)^2}\left(\frac{1}{\epsilon}+\frac{3}{2}+\log\frac{\Lambda^2}{m_u^2+g_\Delta^2\Delta_{us}^2}\right)\right.\nonumber\\
&& -\frac{m_u^4}{(4\pi)^2}\left(\frac{1}{\epsilon}+\frac{3}{2}+\log\frac{\Lambda^2}{m_u^2}\right) + \frac{4(m_u-m_d)^2g_\Delta^2\Delta_{ud}^2}{(4\pi)^2}\left(\frac{1}{\epsilon}+\log\frac{\Lambda^2}{g_\Delta^2\Delta_{ud}^2}\right) - \frac{16\bar{\mu}_{ud}^2g_\Delta^2\Delta_{ud}^2}{(4\pi)^2}\left(\frac{1}{\epsilon}+\log\frac{\Lambda^2}{g_\Delta^2\Delta_{ud}^2}\right)\nonumber\\
&& \left.+\frac{4g_\Delta^4(\Delta_{ud}^2\Delta_{us}^2 + \Delta_{ud}^2\Delta_{ds}^2+\Delta_{us}^2\Delta_{ds}^2)}{(4\pi)^2}\left(\frac{1}{\epsilon}+\log\frac{\Lambda^2}{M^2}\right)\right\}\;,
\end{eqnarray}
\end{widetext}
with cyclic permutation over flavor in the first line and cyclic permutation over flavor and color in the second line.
The counterterms are generated by the tree-level thermodynamic potential as before.
Except for $\delta c_{\ms}$, the other counterterms in the scalar sector
are the same as in the two-flavor case~\cite{kumari,threeflavors}. It reads
\begin{eqnarray}
\delta c_{\ms}&=&{9g^2c\Lambda^{-3\epsilon}\over2(4\pi)^2\epsilon}\;,
\end{eqnarray}
and $c_{\ms}$ runs as
\begin{eqnarray}
    c_{\ms}(\Lambda)&=&{c_0\over[1-{3g_0^2\over(4\pi)^2}\log{\Lambda^2\over\Lambda_0^2}]^{\frac{3}{2}}}\;.
\end{eqnarray}
The new counterterms in the diquark sector are
\begin{eqnarray}
    \delta\lambda_3 &=& \frac{\left[3g^2\lambda_3+4g_\Delta^2\lambda_3-8g^2g_\Delta^2\right]\Lambda^{-4\epsilon}}{(4\pi)^2\epsilon}\;,
    \\
   \delta\lambda_4 &=& \frac{\left[3g^2\lambda_4+4g_\Delta^2\lambda_4+8g^2g_\Delta^2\right]\Lambda^{-4\epsilon}}{(4\pi)^2\epsilon}\;,\\
    \delta\lambda_5 &=& \frac{\left[3\lambda_5g^2+4\lambda_5g_\Delta^2-4g^2g_{\Delta}^{2}\right]\Lambda^{-4\epsilon}}{(4\pi)^2\epsilon}
\\
    \delta\lambda^{\Delta}_1 &=& \frac{4g_\Delta^2\left[2\lambda_{\Delta,1}-g_\Delta^2\right]\Lambda^{-4\epsilon}}{(4\pi)^2\epsilon}\\
    \delta\lambda^{\Delta}_2 &=& \frac{4g_\Delta^2\left[2\lambda_{\Delta,2}-g_\Delta^2\right]\Lambda^{-4\epsilon}}{(4\pi)^2\epsilon}
\\
    \delta\lambda^{\Delta}_3 &=& \frac{8g_\Delta^2\lambda_{\Delta,3}\Lambda^{-4\epsilon}}{(4\pi)^2\epsilon}
\end{eqnarray}
The new running scalar-diquark parameters are
\begin{eqnarray}
    \lambda_{3,\rm\overline{MS}}(\Lambda) &=& \frac{\lambda_{3,0}-\frac{8g_0^2g_{\Delta,0}^2}{(4\pi)^2}\log\frac{\Lambda^2}{\Lambda_0^2}}{\left[1-\frac{3g_{0}^2}{(4\pi)^2}\log\frac{\Lambda^2}{\Lambda_0^2}\right]\left[1-\frac{4g_{\Delta,0}^2}{(4\pi)^2}\log\frac{\Lambda^2}{\Lambda_0^2}\right]}\\
    \lambda_{4,\rm \overline{MS}}(\Lambda) &=& \frac{\lambda_{4,0}+\frac{8g_0^2g_{\Delta,0}^2}{(4\pi)^2}\log\frac{\Lambda^2}{\Lambda_0^2}}{\left[1-\frac{3g_{0}^2}{(4\pi)^2}\log\frac{\Lambda^2}{\Lambda_0^2}\right]\left[1-\frac{4g_{\Delta,0}^2}{(4\pi)^2}\log\frac{\Lambda^2}{\Lambda_0^2}\right]}\;,\\
    \lambda_{5,\rm \overline{MS}}(\Lambda) &=& \frac{\lambda_{5,0}-\frac{4g_0^2g_{\Delta,0}^2}{(4\pi)^2}\log\frac{\Lambda^2}{\Lambda_0^2}}{\left[1-\frac{3g_{0}^2}{(4\pi)^2}\log\frac{\Lambda^2}{\Lambda_0^2}\right]\left[1-\frac{4g_{\Delta,0}^2}{(4\pi)^2}\log\frac{\Lambda^2}{\Lambda_0^2}\right]}\;
\end{eqnarray}
While the new running diquark parameters are
\begin{eqnarray}
        \lambda^{\Delta}_{1,\rm\overline{MS}}(\Lambda) &=& \frac{\lambda_{\Delta1,0}-\frac{4g_{\Delta,0}^4}{(4\pi)^2}\log\frac{\Lambda^2}{\Lambda_0^2}}{\left[1-\frac{4g_{\Delta,0}^2}{(4\pi)^2}\log\frac{\Lambda^2}{\Lambda_0^2}\right]^2}\;,\\
    \lambda^{\Delta}_{2,\rm \overline{MS}}(\Lambda) &=& \frac{\lambda_{\Delta2,0}-\frac{4g_{\Delta,0}^4}{(4\pi)^2}\log\frac{\Lambda^2}{\Lambda_0^2}}{\left[1-\frac{4g_{\Delta,0}^2}{(4\pi)^2}\log\frac{\Lambda^2}{\Lambda_0^2}\right]^2}\;.\\
    \lambda^{\Delta}_{3,\rm \overline{MS}}(\Lambda) &=& \frac{\lambda_{\Delta3,0}}{\left[1-\frac{4g_{\Delta,0}^2}{(4\pi)^2}\log\frac{\Lambda^2}{\Lambda_0^2}\right]^2}\;.
\end{eqnarray}\newpage
\noindent The final renormalized thermodynamic potential in the CFL phase is
\begin{widetext}
\begin{eqnarray}
    \Omega^{\rm CFL}_{0+1} &=&  \Omega_0^{\rm CFL}+\frac{g_0^4\phi_u^4}{16(4\pi)^2}\left( 2\log\frac{\Lambda_0^2}{m_u^2+g_{\Delta,0}^2\Delta_{ud,0}^2} + 2\log\frac{\Lambda_0^2}{m_u^2+g_{\Delta,0}^2\Delta_{us,0}^2} -\log\frac{\Lambda_0^2}{m_u^2}+ \frac{9}{2}\right)+  d\rightarrow s  + u\rightarrow s\nonumber\\
    && - \frac{16g_{\Delta,0}^2\bar{\mu}_{ud}^2\Delta_{ud,0}^2}{(4\pi)^2}\log\frac{\Lambda_0^2}{g_{\Delta,0}^2\Delta_{ud,0}^2} + \frac{2g_{\Delta,0}^4\Delta_{ud,0}^4}{(4\pi)^2}\left(\log\frac{\Lambda_0^2}{m_u^2+g_{\Delta,0}^2\Delta_{ud,0}^2}+\log\frac{\Lambda_0^2}{m_d^2+g_{\Delta,0}^2\Delta_{ud,0}^2}+3\right) + d\rightarrow s + u\rightarrow s\nonumber\\
    && +\frac{g_0^2g_{\Delta,0}^2(\phi_{u,0}-\phi_{d,0})^2\Delta_{ud,0}^2}{(4\pi)^2}\log\frac{\Lambda_0^2}{g_{\Delta,0}^2\Delta_{ud,0}^2}  +  d\rightarrow s + u\rightarrow s\nonumber\\
    && +\frac{g_0^2g_{\Delta,0}^2\Delta_{ud,0}^2}{(4\pi)^2}\left(\phi_{u,0}^2\left[\log\frac{\Lambda_0^2}{m_u^2+g_{\Delta,0}^2\Delta_{ud,0}^2}+\frac{3}{2}\right] + \phi_{d,0}^2\left[\log\frac{\Lambda_0^2}{m_d^2+g_{\Delta,0}^2\Delta_{ud,0}^2}+\frac{3}{2}\right]\right) + d\rightarrow s +u\rightarrow s\nonumber\\
    && + \frac{4g_{\Delta,0}^4(\Delta_{ud,0}^2\Delta_{us,0}^2 + \Delta_{ud,0}^2\Delta_{ds,0}^2+\Delta_{us,0}^2\Delta_{ds,0}^2)}{(4\pi)^2}\log\frac{\Lambda_0^2}{M^2} + \Omega_{1}^{\rm CFL, fin}\;,
\end{eqnarray}
\end{widetext}
where there is an additional factor of \(\sqrt{2}\) in front of any \(\phi_s\) under  permutation.


Although we postpone numerical work to a follow-up paper, it will still be instructive
with an application that shows the usefulness of our approach.
In unpaired quark matter with three massless flavors, the system is neutral with respect
to all charges $Q$, $Q_3$, and $Q_8$ for $\mu_e=\mu_3=\mu_8=0$.
since the trace of the three generators is zero~\footnote{For two massless flavors, electric charge neutrality requires a nonzero background electron density.}

In the CFL phase with three degenerate quark flavors, the system is charge neutral
if all quark flavors have a common chemical potential, which we denote by $\bar{\mu}$.
In this case, the nine dispersion relations for the quarks can be found explicitly. 
They split up into an octet with a gap $\Delta$ and a singlet with a gap $2\Delta$,
\begin{eqnarray}
    E_{\Delta^{{\pm}}\rm}^{\rm o} &=& \sqrt{(E_{\pm})^2+g_\Delta^2\Delta^2}
    \;, \\
    E_{\Delta^{{\pm}}}^{\rm s} &=& \sqrt{(E_{\pm})^2+4g_\Delta^2\Delta^2}
    \;,    
\end{eqnarray}
where
\begin{eqnarray}
    E_{\pm} &=& \sqrt{p^2+m_c^2}\pm\bar{\mu}\;,
\end{eqnarray}
where $m_c={1\over\sqrt{6}}g\bar{\sigma}_0\equiv g\phi$ is the common quark mass.
The contribution to the thermodynamic potential for a quark with energy $E$ and
chemical potential is well known,
\begin{eqnarray}
\Omega_1^{}&=&\Lambda^{-2\epsilon}\int_p\left[E+(E-\mu)\theta(E-\mu)\right]\;.    
\end{eqnarray}
In the present case, the first term is ultraviolet divergent and the second term is finite. The subtraction terms are constructed as described earlier. We can then write
\begin{widetext}
\begin{eqnarray}
\nonumber
\Omega_1^{\rm CFL, div}&=&-\Lambda^{-2\epsilon}\int_p\left[
16\sqrt{p^2+g^2\phi^2+g_{\Delta}^2\Delta^2}+2\sqrt{p^2+g^2\phi^2+4g_{\Delta}^2\Delta^2}
+{8g_{\Delta}^2\bar{\mu}^2\Delta^2\over(p^2+g^2\phi^2+g_{\Delta}^2\Delta^2)^{3\over2}}
\right.\\ &&\left.
+{4g_{\Delta}^2\bar{\mu}^2\Delta^2\over(p^2+g^2\phi^2+4g_{\Delta}^2\Delta^2)^{3\over2}}
\right]\;.
\\
\label{finito}
\Omega_1^{\rm CFL, fin}&=&\Omega_1^{\rm CFL}-\Omega_1^{\rm CFL, div}\;.
\end{eqnarray}
The finite expression in Eq.~(\ref{finito}) can be evaluated numerically directly 
in  $d=3$ dimensions. 
\end{widetext}
We are in particular interested of the behavior of the system at large chemical potential, i.e deep into the CFL phase. For sufficiently large $\bar{\mu}$, we can
ignore the quark masses.
In the limit of zero quark masses, the subtraction term $\Omega_1^{\rm fin}$ 
reduces to contribution from $N_cN_f$ massless quarks.
The renormalized thermodynamic potential then simplifies to
\begin{eqnarray}
\nonumber
\Omega_{0+1}^{\rm CFL}&=&3(m_{\Delta,0}^2-4\bar{\mu}^2)\Delta_0^2+\frac{3}{2}\left[3\lambda_{1,0}^{\Delta}+\lambda_{2,0}^{\Delta}+\frac{3}{2}\lambda_{3,0}^\Delta\right]\Delta_0^4
\\ &&
\nonumber
-{12\bar{\mu}^4\over(4\pi)^2} + \frac{36g_{\Delta,0}^4\Delta_0^4}{(4\pi)^2}
\\ &&
\nonumber
-{16g_{\Delta,0}^2\bar{\mu}^2\Delta_0^2\over(4\pi)^2}\left[2\log{\Lambda_0^2\over g_{\Delta,0}^2\Delta_0^2}+\log{\Lambda_0^2\over4g_{\Delta,0}^2\Delta_0^2}\right]
\\&&
+{8g_{\Delta,0}^4\Delta_0^4\over(4\pi)^2}\left[\log{\Lambda_0^2\over g_{\Delta,0}^2\Delta_0^2}+2\log{\Lambda_0^2\over4g_{\Delta,0}^2\Delta_0^2}\right]
\;.
\end{eqnarray}
The thermodynamic potential is of the same form as derived in the two-flavor case
in Ref.~\cite{us}. In the limit, where $\bar{\mu}$ is much larger than the gap, one can 
easily find the solution to the gap equation
\begin{eqnarray}
    \Delta_0 &=& \frac{\Lambda_0}{g_{\Delta,0}}2^{-\frac{1}{3}}e^{\frac{(4\pi)^2}{8g_{\Delta,0}^2}-\frac{1}{2}} = 2^{-\frac{1}{3}}\Delta_0^{\rm 2SC}\;,
\end{eqnarray}
where $\Delta_0^{\rm 2SC}$ is the gap in the 2SC phase. The relation between the
octet gap and the gap in the 2SC phase
is a perturbative-QCD result at asymptotic high densities as first
shown in Ref.~\cite{gaprel}.
Using the expression for the gap, we obtain simple expressions for
the pressure and energy density 
\begin{eqnarray}
p&=&
{12\bar{\mu}^4\over(4\pi)^2}
+{48g_{\Delta,0}^2\bar{\mu}^2{\Delta}_0^2\over(4\pi)^2}
\;,\\
\epsilon&=&
{36\bar{\mu}^4\over(4\pi)^2}
+{48g_{\Delta,0}^2\bar{\mu}^2{\Delta}^2_0\over(4\pi)^2}\;.
\end{eqnarray}
Up to corrections due to the finite mass of the $s$-quark, the result for the pressure
and energy density are the same as those obtained using the NJL model~\cite{rajagopal1,steinerneutral}.
From these expressions, we finally obtain the speed of sound
\begin{eqnarray}
c_s^2&=&{1\over3}\left[1+{4\over3}{g_{\Delta}^2\Delta_0^2\over\bar{\mu}^2}\right]\;,    
\end{eqnarray}
which shows that $c_s^2$ approaches the conformal limit from above.
This is a generic feature of the model: The same behavior was found in the case of finite isospin chemical potential $\mu_I$ and pion condensation~\cite{isokojo,farias,minato,brandtqm,us2}.

\section{Summary}
In the present paper, we have extended the two-flavor QMD model of Ref.~\cite{us} to include the new degrees of freedom $\eta$ and $\vec{a}$.
This allows for isospin breaking effects ($\mu_u\neq m_d$) either in the vacuum,
or in a medium due to finite $\mu_I$. This is in principle relevant in the pion-condensed
phase and in color-superconducting phases.
We have renormalized the thermodynamic potential in the 2SC phase for two flavors
and in the CFL phase for three flavors, and determined the counterterms of the
couplings involving the diquark degrees of freedom.
in the process. The running masses and couplings satisfy a set of (coupled) differential
equations that were solved. Using the results for the running parameters, we obtained
a renormalization-group improved thermodynamic potential.

The thermodynamic potential was calculated including the effects of quark loops.
To be consistent with this approximation, one must determine the relation between
the bare and renormalized parameters in the Lagrangian in the same approximation.
In Ref.~\cite{onshell}, it was shown how one can consistently determine the parameters
of the two-flavor quark-meson model by combining the on-shell and $\overline{\rm MS}$ schemes by expressing them in terms of the meson masses and the pion decay constant.
Later, these complicated relations were derived in the extended models considered here~\cite{kumari,threeflavors}.

In the original formulation of the two-flavor QMD model, four new unknown parameters,
$m_{\Delta}^2$, $g_{\Delta}^2$, $\lambda_3$, and $\lambda_{\Delta}$, were introduced.
The extension discussed here requires an additional coupling $\lambda_{\rm det}$.
In the three-flavor case,  a total of seven new operators with seven unknown parameters
involving diquarks are required. Based on previous work~\cite{us,us2}, we anticipate that the results will mainly be sensitive to the values of $m_{\Delta}^2$ and $g_{\Delta}^2$. The three-flavor QMD model is a possible starting point to describe quark phases in hybrid stars. This requires that charge neutrality
for the three commuting charges $Q$, $Q_3$, and $Q_8$ is imposed so that the 
pressure and energy density are expressed in terms of $\mu_B$ alone.
The prediction of macroscopic quantities such as masses, radii, and
tidal deformation will depend on the parameters of the model.
One way to determine the parameters is to confront the model with astrophysical
observations in conjunction with Bayesian analysis.

\appendix

\bibliographystyle{apsrmp4-1}

\end{document}